\documentclass[]{article}
\usepackage{graphics}
\usepackage{graphicx}

\begin{document}

\newcommand{\be}{\begin{equation}}
\newcommand{\ee}[1]{\label{#1} \end{equation}}
\newcommand{\ba}{\begin{eqnarray}}
\newcommand{\ea}[1]{\label{#1} \end{eqnarray}}
\newcommand{\nl}{\nonumber \\}
\newcommand{\iint}{\int_0^{\infty}\!}
\newcommand{\pd}[2]{\frac{\partial #1}{\partial #2}}
\newcommand{\vs}{\vspace{3mm}}

\renewcommand{\d}[1]{{\rm d}#1}

\title{{\bf Microscopic Origin of Non-Gaussian Distributions of Financial Returns}}

\author{
 {\sc T.~S.~Bir\'o}$^1$ and {\sc R.~Rosenfeld}$^2$ \\[1em]
$^1$ KFKI RMKI, Budapest, Hungary \\[0em]
$^2$ Instituto de F\'{\i}sica Te\'orica \\ State University of S\~ao Paulo, S\~ao Paulo, Brazil
}

\date{\today}

\maketitle

\begin{abstract}
In this paper we study the possible microscopic
origin of heavy-tailed probability density distributions
for the price variation of financial instruments.
We extend the standard log-normal process to include
another random component in the so-called stochastic volatility
models. We study these models under an assumption, akin to the Born-Oppenheimer approximation,
in which the volatility
has already relaxed to its equilibrium distribution and acts 
as a background to the evolution of the price process.  
In this approximation, we show that all models of stochastic volatility should
exhibit a scaling relation in the time lag of zero-drift modified log-returns. We verify that the Dow-Jones Industrial
Average index indeed follows this scaling.
We then focus on two popular stochastic volatility models, the Heston and Hull-White models.
In particular, we show that in the Hull-White model the resulting probability distribution of log-returns in
this approximation corresponds to the Tsallis (t-Student) distribution. The Tsallis parameters are
given in terms of the microscopic stochastic volatility model.
Finally, we show that the log-returns for 30 years Dow Jones index data is well fitted by a Tsallis distribution,
obtaining the relevant parameters.
\end{abstract}

\section{Introduction}

As any mature field, Finance has adopted a simple model developed 
over the years that attempts to describe
the behaviour of random time fluctuations in the prices of commodities 
or stocks observed in the markets \cite{Hull}.
This model, which one could call the Standard Model of Finance (SMF), in spite 
of its many shortcomings has established
a common language in a specific framework that immediately allows for generalizations. 
It predicts, for instance, prices
for contracts on stocks, usually called derivative contracts \cite{BlackScholes}. 
The SMF assumes that the fluctuations of the stock prices follow a log-normal probability 
distribution function. It was first suggested by Osborne in 1959 \cite{Osborne} and 
independently by Samuelson \cite{taqqu}, improving on an earlier model by guaranteeing 
non-negative stock prices.
 
In recent years, the large amount of financial data available has prompted many empirical 
investigations of the probability distributions of returns, defined as the logarithm of 
the ratio of stock prices separated by a given time lag. 
The simple log-normal assumption of the SFM would predict a Gaussian distribution for 
the returns with variance growing linearly with the time lag. What is actually found 
is that the probability distribution for high frequency data
usually deviates from normality, presenting heavy tails. 
Fits have been made using the truncated L\'evy distribution \cite{Stanley} and the 
so-called Tsallis (or t-Student) distribution \cite{Tsallis}.

An important question that we want to address is the nature of the microscopic stochastic 
process that may lead to these non-Gaussian distributions. 
For example, Borland proposed a feedback process where the diffusion coefficient is 
related to the macroscopic probability distribution, resulting in a non-linear 
Fokker-Planck equation whose solution is a time-dependent Tsallis
distribution \cite{Borland}.

In this paper we study an extension of the SFM in which the diffusion of the 
stock price is itself represented by another independent stochastic process.
In a very general way, one can divide these models into 
continuous- time stochastic volatility models and discrete-time Generalized Autoregressive
Conditional Heteroskedasticity (GARCH) models \cite{HestonNandi}. 
Both classes can in principle describe
some stylized facts such as volatility clustering \cite{stylized}.
The second class of models, used in standard econometric analysis, is based on 
time series data where previous values of the volatility and stock prices are used
to calculate subsequent values. 

In the following we will assume that volatility is not directly observed, and hence we will work in the 
context of the first class of models, where volatility can be interpreted as a 
hidden Markov process \cite{hidden}.   
We show that in these stochastic volatility models there is a useful
assumption that can be made and which results in probability distribution of returns with heavy tails 
and a simple scaling time dependence. In particular, we find that the Tsallis 
distribution follows from the Hull-White stochastic volatility model in this case.
We test this assumption with daily data from the Dow Jones Industrial Average 
index and find a better agreement than the SFM.   

\vs
\section{Coupled stochastic processes for log-return and volatility}

The price of a stock, $s(t)$, is a stochastic variable on the top of a deterministic
exponential growth (with inflation rate $\mu>0$). It is therefore
customary to regard the modified log-return,
\be
 x = \ln \frac{s_t}{s_0} - \mu t,
\ee{LOG_RETURN_DEF}
as the indicator variable. It follows a Langevin equation of type
\be
 \d{x} = -a(v)\d{t} + \sqrt{v} dW_{1}
\ee{X_LANGEVIN}
with $dW_1$ being a Wiener process with zero mean and variance $dt$ and
$\sigma = \sqrt{v}$ is called the volatility.
The function $a(v)$ reflects the prescription when deriving the $x$-process
with additive noise from the $s$-process with multiplicative noise.
In the Ito-scheme it is $a(v)=v/2$. There exist, however, one particular
Stratonovich scheme where it vanishes $a(v)=0$. Another possibility 
is to modify the equation (\ref{LOG_RETURN_DEF}) by adding
a factor $1/2 \int dt \; v(t)$.

Equation (\ref{X_LANGEVIN}) defines the SFM.
An extension of the SFM considers the possibility that the variance $v=\sigma^2$ 
itself is governed by stochastic effects, as suggested by
phenomenological observations on financial markets \cite{Mantegna}.
The proposed models have a first order deterministic part
causing an exponential approach to the mean volatility, $\theta$
and a noise term possibly influenced by the volatility itself:
\be
 \d{v} = -\gamma(v-\theta)\d{t} + b(v) dW_2.
\ee{V_LANGEVIN}
The second Wiener process, $dW_2$, may or may not be correlated with
the first one. 
We will consider two main models of stochastic volatility developed
in the finance literature, namely the Heston and the Hull-White models.
The Heston model uses $b(v)=\kappa \sqrt{v}$ \cite{Heston} and in the
Hull-White model $b(v)=\kappa v$ \cite{HW}.

Our purpose is to investigate approximations which enable a simplified
treatment predicting the distribution of $x$ (or $s$) as a function of
time. 

\vs
\section{Born-Oppenheimer approximation}

Assuming that one of the coupled dynamical stochastic variables ($v$) reaches
its stationary distribution and then acts for the dynamics of the other ($x$)
as an instantaneous, time-independent background is  analog to the Born-Oppenheimer (BO) 
approximation applied successfully in solid state and atomic physics. We will make this assumption 
in the remaining of the paper, explore its consequences and test it against data.
The $v$ -process can be solved in itself; the corresponding Fokker-Planck equation
is given as
\be
 \pd{\Pi}{t} + \pd{}{v} \, \gamma(v-\theta)\Pi 
 + \frac{1}{2} \pd{^2}{v^2} b^2(v)\Pi = 0.
\ee{V_FOKKER_PLANCK}
The stationary detailed balance solution satisfies
\be
 \pd{}{v} \,  b^2(v) \Pi  = - {2\gamma(v-\theta)} \Pi.
\ee{V_BALANCE}
The solution, the balance distribution of $v$ is given by
\be
 \Pi(v) = \frac{N}{b^2(v)} \, \exp\left(-2\gamma\int^{v} \frac{v'-\theta}{b^2(v')} dv' \right)
\ee{PI_V}

According to the BO-approximation we average over this stationary balance 
probability of $v$ the solution of the diffusion process of eq.(\ref{X_LANGEVIN}) at a given $v$,
\be
 P(x;v,t) = \frac{1}{\sqrt{2\pi v t}} \exp\left(-\frac{(x+a(v)t)^2}{2vt}\right).
\ee{X_DIFFUSION}
We obtain an approximation for the time evolution of the log-return probability:
\be
 P_t(x) = \int_0^{\infty} \!\!dv \, \Pi(v) \, P(x;v,t).
\ee{X_PROB}
Collecting all terms together we arrive at
\be
 P_t(x) = \frac{N}{\sqrt{2\pi t}} \int_0^{\infty} \!\!dv \, \frac{1}{b^2(v)\sqrt{v}} \,
 \exp\left(-\frac{(x+a(v)t)^2}{2vt} -2\gamma\int \frac{v'-\theta}{b^2(v')} dv' \right). 
\ee{X_GENERAL_PROB}

\vs
\section{Scaling in the BO approximation}

In the case of a compensated Ito-term, i.e. $a(v)=0$, the dependence on the time lag $t$
in eq. (\ref{X_GENERAL_PROB})
is via the combination $x^2/2t$ and a $1/\sqrt{t}$ in the normalization:
\be
 P_t(x) = \frac{N}{\sqrt{t}} \, \rho \left(\frac{x^2}{2t} \right).
\ee{PxSCALED}

The scaling hypothesis (\ref{PxSCALED}) can be checked by detrending and normalizing the data.
First one considers the log-return with a unit time-lag as
\be
 \xi_i = \ln \frac{s_{i+1}}{s_i}.
\ee{X_SERIES}
These values usually show a trend with the index $i$ for a fixed $t$.
We subtract the best fitted linear trend by the Gaussian least square method.
From
\be
 \frac{1}{2} \sum_i \left(\xi_i-a-b*i \right)^2 = {{\rm min}} 
\ee{GAUSS}
it follows by derivation with respect to the fit parameters $a$ and $b$:
\ba
 \sum_i (\xi_i-a-b*i)   &=& 0, \nl
 \sum_i (\xi_i-a-b*i)*i &=& 0.
\ea{FIT_PARAMS}
From this we obtain 
\ba
 b = \frac{6}{N-1} \left(\overline{i\xi_i}-\overline{\xi_i}\right) \nl
 a = \overline{\xi_i} - b \frac{N+1}{2}
\ea{FIT_SOLUTION}
with
\ba
 \overline{\xi_i} = \frac{\sum_i \xi_i}{\sum_i 1} = \frac{1}{N} \sum_i \xi_i, \nl
 \overline{i\xi_i} = \frac{\sum_i i\xi_i}{\sum_i i} = \frac{2}{N(N+1)} \sum_i i\xi_i.
\ea{AVERAGES}
Next we compose the subtracted data as
\be
 y_i = \xi_i - a - b*i.
\ee{SUBDATA}
These have by construction zero mean, ($\sum_i y_i=0 $) and 
zero index expectation value ($\sum_i i y_i = 0$). 
Finally the variance is normalized to one by considering
\be
 x_i = \frac{y_i\sqrt{N}}{\sqrt{\sum_k y_k^2}}.
\ee{DETREND_NORMAL}
The $x_i$ we call (linearly) detrended and normalized log return data.


The distribution of these data is obtained by binning the $x_i$ values:
the distance between the minimal and maximal $x_i$ is divided into $B$ intervals,
so that $X_k=x_{{\rm min}}+k(x_{{\rm max}}-x_{{\rm min}})/B$ is the beginning
of the $k-th$ interval. Whenever a given $x_i$ is between $X_k$ and $X_{k+1}$
the counter $P_k$ is increased by one. The normalization $\sum_k P_k=N$
is checked and $\sum_k kP_k$ is calculated. The normalized distribution 
is reconstructed as $P_k/N=P(X_{k+1/2})$. We have tested the binning program with
hoax data constructed from a normal Wiener process. 

\begin{figure}
\includegraphics[width=0.7\textwidth,angle=-90]{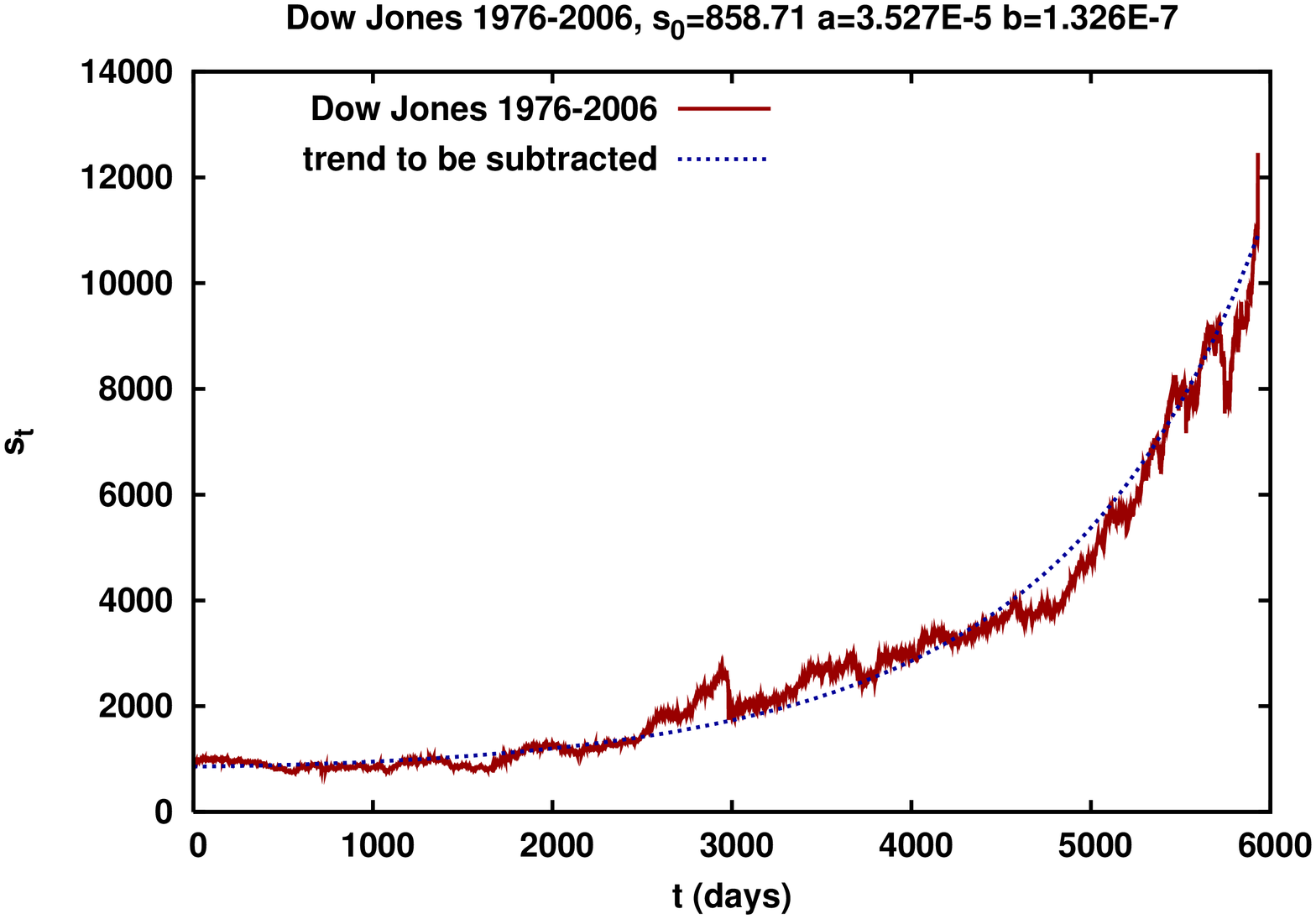}
\includegraphics[width=0.7\textwidth,angle=-90]{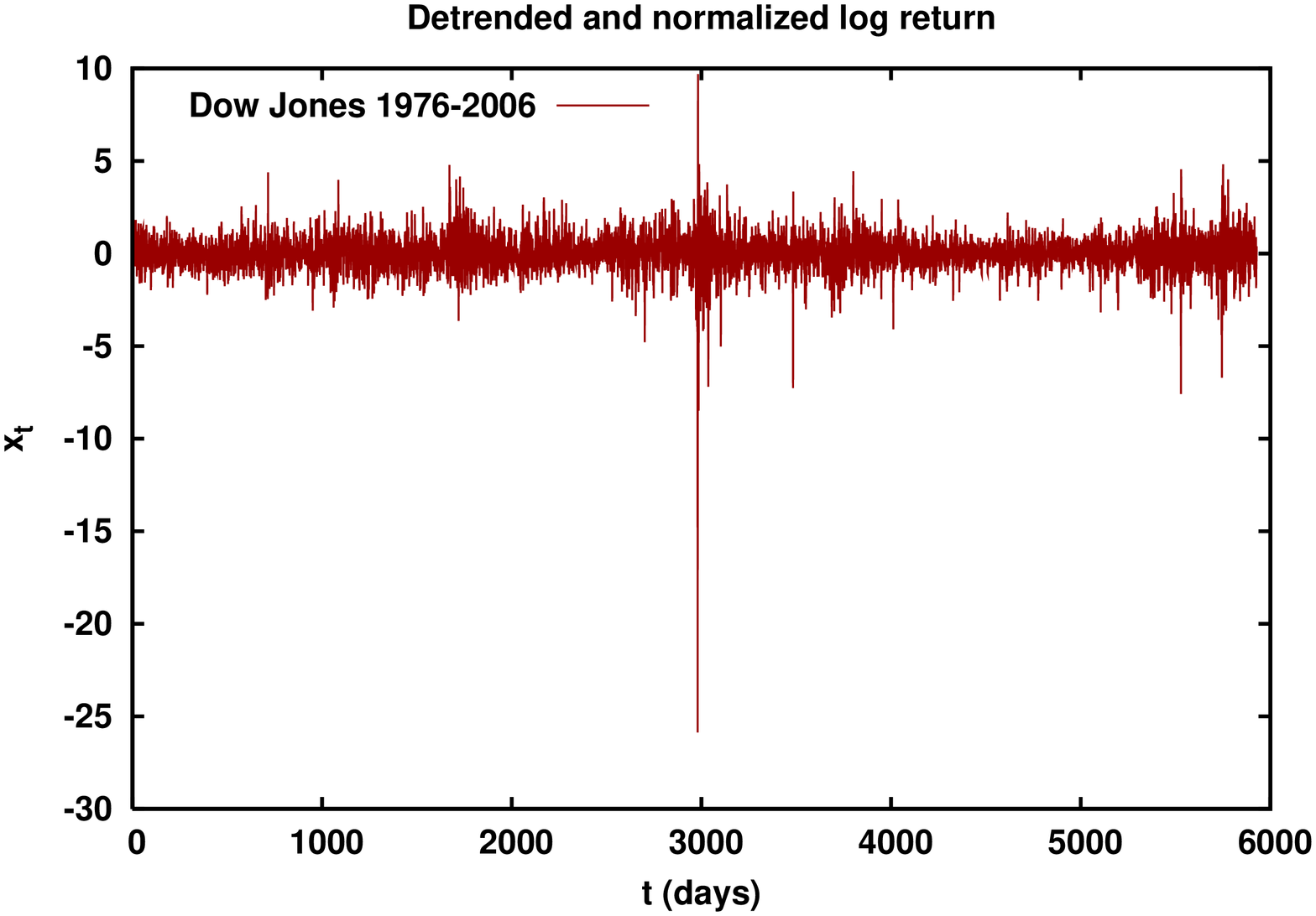}
\caption{\label{FIG_TREND}
 Time data of Dow Jones daily closing average $s_t$, showing the trend to be subtracted 
in our procedure (top panel) and the detrended and normalized
 log return $x_t$ (bottom panel) for 30 year data.
}
\end{figure}

In order to check the scaling hypothesis we use the daily closing values of the Dow Jones Industrial Average (DJIA) index
from January 1976 to December 2006, with $N=5930$ data points and we adopt $B=100$ bins.
In figure \ref{FIG_TREND} we show the index time series and the detrended and normalized returns.


\begin{figure}
\includegraphics[width=0.7\textwidth,angle=-90]{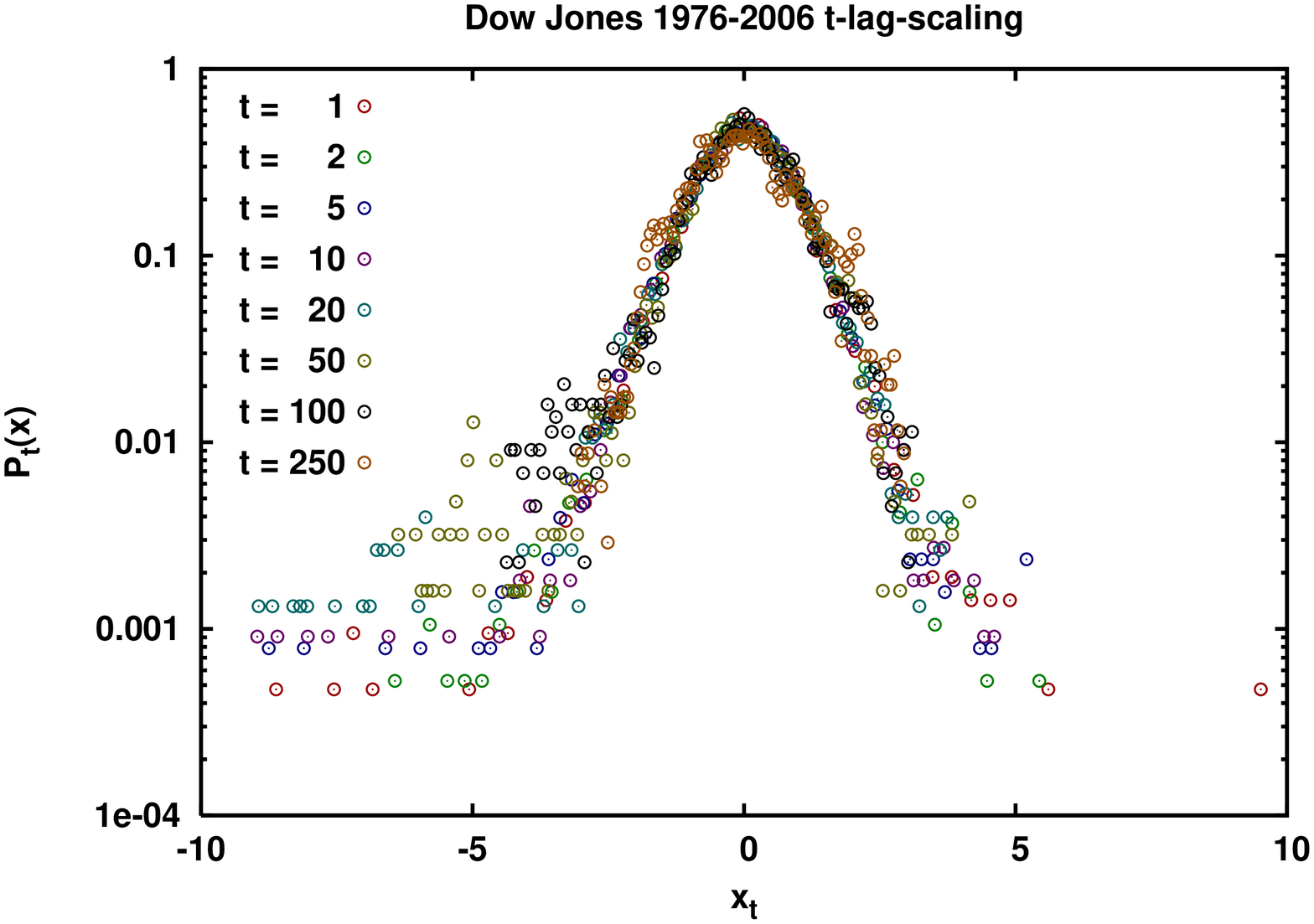}
\caption{\label{DJ30_ALL}
 The distribution of detrended and normalized log-returns $x_i(t)$ with 
  different time lags $t$ for Dow Jones data 1996-2006. 
 The near coincidence of points indicates scaling and $a(v)\approx 0$.
}
\end{figure}

Figure \ref{DJ30_ALL} presents the detrended and normalized log-returns 
at different time-lags from 1 day to 250 days. The closeness of these data
indicate that the scaling (detrending) is successful.

The universal scaling for $a(v)=0$ case is (cf. eq.\ref{PxSCALED})
\be
P_t(x)= \frac{1}{\sqrt{t}} \, f\left(\frac{x}{\sqrt{t}} \right)
\ee{UNIVERSAL}
for any time-lag $t$. 
This is testable on the detrended data with different time-lags.


%

\begin{figure}
\includegraphics[width=0.7\textwidth,angle=-90]{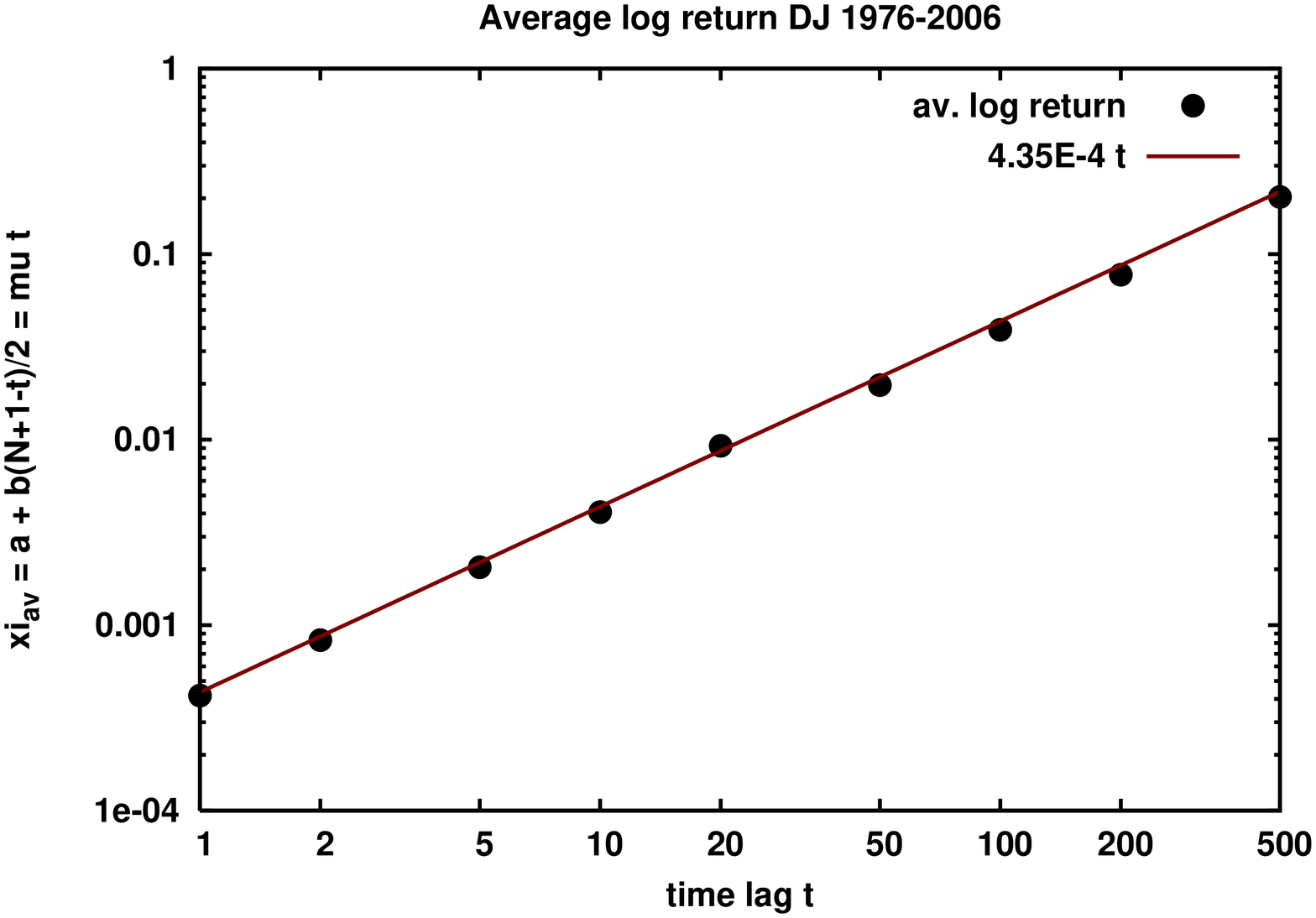}
\includegraphics[width=0.7\textwidth,angle=-90]{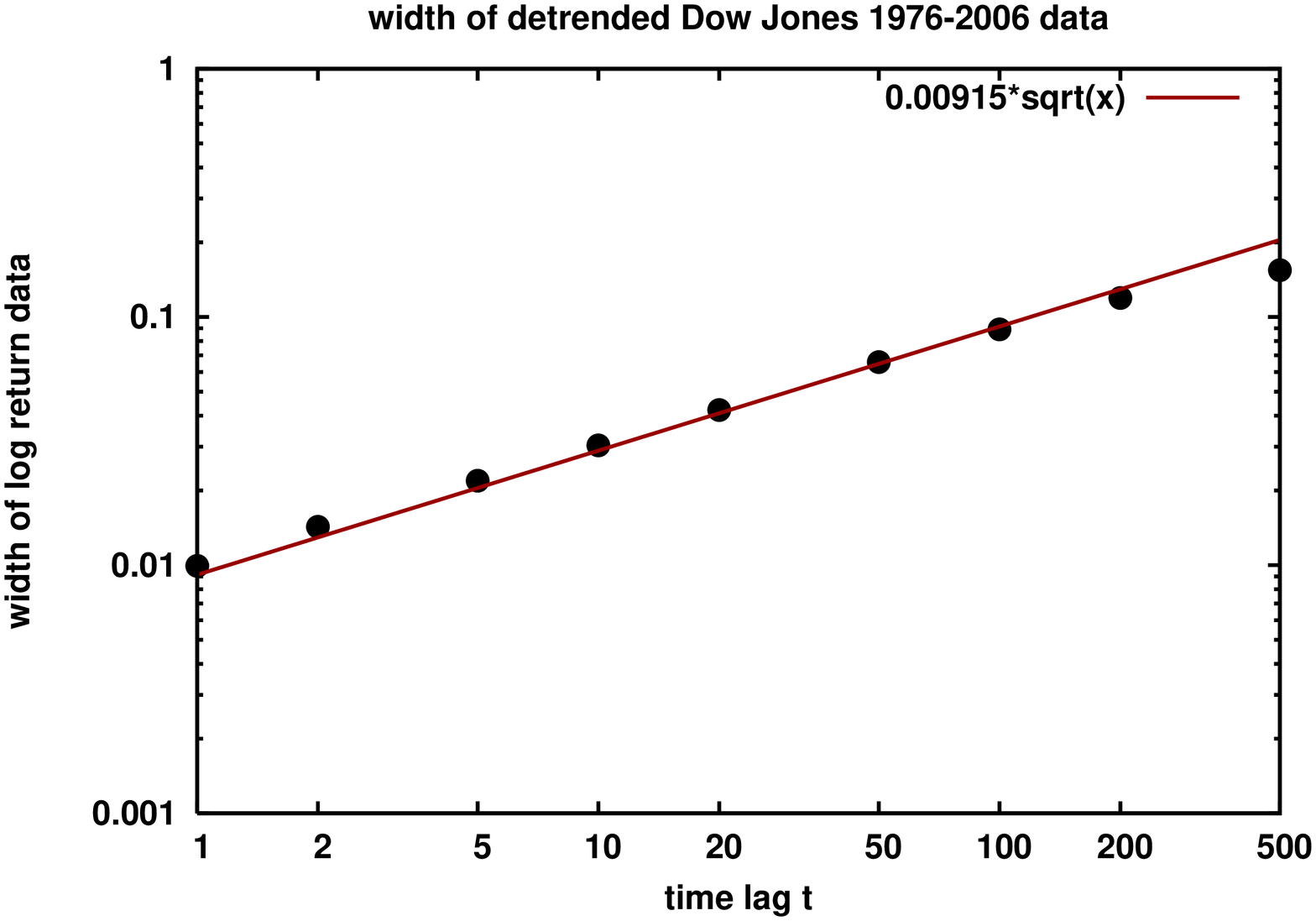}
\caption{\label{DJ_TREND}
 The trend in log-returns $\xi_i(t)$ with time lag $t$ for Dow Jones
data 1996-2006 (top) and the width of the detrended data showing
a scaling compatible with the assumption $a(v)=0$ in the stochastic 
log return process (bottom).
}
\end{figure}

In Figure \ref{DJ_TREND} the log returns $\xi_i(t) = \ln(s_{i+t}/s_i)$ for a given time lag $t$
show an average trend. The $a_t$ and $b_t$ parameters are obtained for several different
time lags (from 1 to 500 days), as well as the width of detrended data before 
normalization. The average of $\xi_i(t)$  on the top follows the straight line
$\overline{\xi}_i(t) = a_t + b_t(N+1-t)/2 = \mu t$ with $\mu \approx 4.35\cdot 10^{-4}$
meaning a yearly $8.5\%$ earnings (an average year had 5930/30=197 working days of the borse).
The bottom picture shows the width of detrended data which is proportional to
the square root of the time-lag on a double logarithmic scale. The fulfillment of
this proportionality indicates that the scaling $P_t(x)=f(x/\sqrt{t})/\sqrt{t}$ is
realized by the data - since $\int P_t(x) dx=1$. This supports the
BO approximation with $a(v)=0$.

\vs
\section{Heston model in BO}

The probability of returns in the Heston model
has been extensively studied \cite{DY}.
In the Heston model, $b(v)=\kappa\sqrt{v}$ and
the stationary distribution is a Gamma distribution:
\be
 \Pi(v) = \frac{\alpha^{\alpha}}{\theta^{\alpha}\Gamma(\alpha)} v^{\alpha-1} e^{-v\alpha/\theta}
\ee{V_STAT}
with $\alpha=2\gamma\theta/\kappa^2$.
In the Ito-scheme $a(v)=v/2$ and the above integral
(\ref{X_GENERAL_PROB}) can be determined analytically. 
The BO approximation corresponds to what
ref. \cite{DY} calls the short time behaviour.
Our basic assumption is that the volatility has already reached 
its stationary distribution.
We obtain
\be
 P_t(x) = \frac{1}{\Gamma(\alpha)} \left(\frac{4\gamma|x|}{\kappa^2 f t} \right)^{\alpha} \sqrt{\frac{f}{\pi|x|}}
 e^{-x/2} K_{\alpha-1/2}\left(\frac{f|x|}{2}\right)
\ee{P_OF_X_HESTON}
with
\be
 f=\sqrt{1+\frac{16 \gamma}{\kappa^2 t}}
\ee{fDEF}
and $K_{\nu}(z)$ Bessel K-function. For large times $f\rightarrow 1$.
For large positive $x$ (extreme large wins on $s$) the Bessel function is dominated by
an exponential, so
\be
 P_t^{(win)}(x) \sim x^n e^{-\frac{f+1}{2} x}
\ee{PLUS_EXP_FACTOR}
and for large negative $x$ (extreme losses) it is dominated by another exponential
\be
 P_t^{(loss)}(x) \sim x^n e^{\frac{f-1}{2}x}.
\ee{MINUS_EXP_FACTOR}
Since $f>1$ at all times it means that in  this model losses tend to show a 
fatter tail (slower decrease) in $x$ than wins. For short times, $t < 16 \gamma/\kappa^2$ this difference reduces.
Note that $e^{-f_{\pm}x}=s^{-f_{\pm}}$ is a power-law behavior in $s$,
denoting by $f_{\pm}=(f\pm 1)/2$.

In the case of $a(v)=0$ this model is also subject to the scaling 
$P_t(x)=\rho(x/\sqrt{t})/\sqrt{t}$
as the normal diffusion. Now the $e^{-x/2}$ factor is not present and 
$f=4\sqrt{\gamma/\kappa^2 t}$ in eq.(\ref{P_OF_X_HESTON}). The distribution
$P_t(x)$ becomes symmetric for wins and losses (relative to the mean trend).

\vs 
\section{Hull-White model in BO}

We now turn our attention to the probability distribution function
resulting from the Hull-White stochastic volatility model in the BO approximation. 
To our knowledge this subject has not been discussed
in the literature. 


In the Hull-White model $b=\kappa v$ and the balanced volatility follows a Gamma-distribution
in the reciprocal variable $y=1/v$, the so-called inverse-Gamma distribution:
\be
 \Pi(y) =  \frac{(\beta\theta)^{\beta+1}}{\Gamma(\beta+1)} \, y^{\beta} \,
 e^{-\beta\theta y}
\ee{INV_PI_Y}
with $\beta=2\gamma/\kappa^2$.
The BO-approximated $x$ distribution in the special scheme with $a(v)=0$ 
becomes:
\be
 P_t(x) = \frac{1}{\sqrt{2\pi t}} \int_0^{\infty}\!\!\!\!\!\!dy  
  \, \, \Pi(y) \sqrt{y} e^{-\frac{x^2}{2t}y}.
\ee{SIMPLE_PX}
Regarding the quantity $E=x^2/2t$ as an abstract ``energy'' the above formula
represents a Boltzmann-Gibbs energy distribution under the influence of fluctuating
``temperature'', which is identified with the variance. 
In particular a Gamma-distributed inverse temperature
is known to lead to a cut power-law (Tsallis, or t-Student) distribution in
the energy variable:
\be
 P_t(x) =  N(\beta,\beta\theta t) \, 
          \left(1+\frac{x^2}{2\beta\theta t} \right)^{-(\beta+3/2)}
\ee{PX_TSALLIS}
with
\be
  N(\beta,z) = \frac{1}{\sqrt{2 z}} \frac{\Gamma(\beta+3/2)}{\Gamma(\beta+1)\Gamma(1/2)}.
\ee{NORM_TSALLIS}
The quotient of $\Gamma$ functions constitutes Bernoulli's Beta-function $B(\beta+1,1/2)$.
A Gaussian distribution with variance given by $\theta t$ is obtained in the $\beta \rightarrow \infty$ 
limit, which can be thought of as the limit of a deterministic volatility ($\kappa \rightarrow 0$).

This result is symmetric in win and loss percentages, i.e. for all positive and
negative $x$ values. As in the Heston model, asymmetries can be introduced
by choosing a non-zero value of $a(v)$.
 On the other hand, it is easy to see that the large $x$ behavior
is now a power-law in $x$, not in $s$. Nevertheless for large powers $\beta$
the cut power law distribution becomes quite close to the exponential
for small and intermediate arguments, so the difference may influence the extreme
large $x$ ($s$) values only.

\begin{figure}
\includegraphics[width=0.7\textwidth,angle=-90]{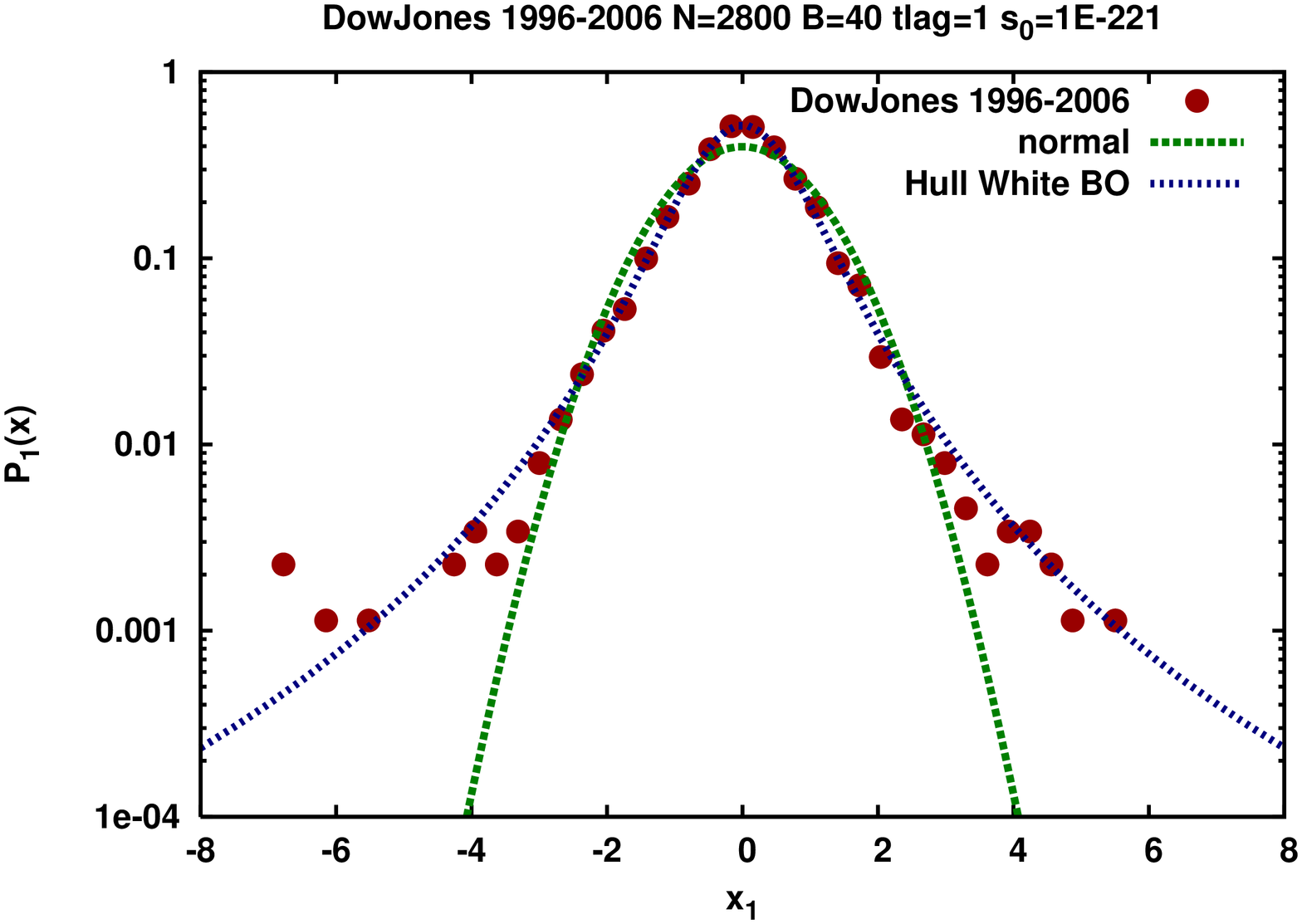}
\includegraphics[width=0.7\textwidth,angle=-90]{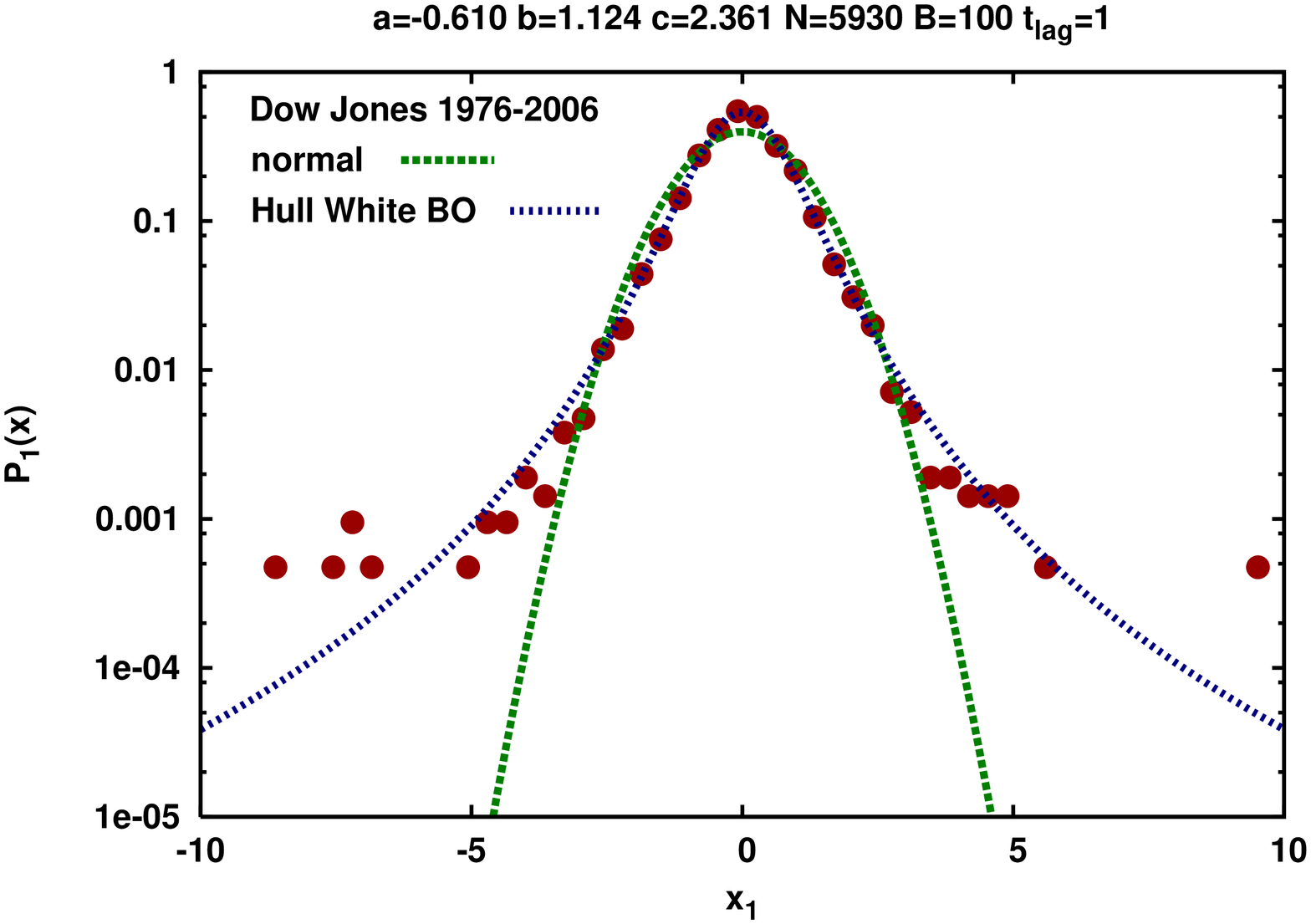}
\caption{\label{FIG_DJFULL}
 The distribution of normalized and detrended log-returns $P_1(x)$ for Dow Jones
data 1996-2006 (top) and 1976-2006 (bottom). The Hull White model fit in BO approximation
is given by $\ln P_1(x) = a - c*\ln(1+bx^2/2)$. This upon eqs.(\ref{PX_TSALLIS},\ref{NORM_TSALLIS}) results in
the values $\beta=0.861$ and $\theta=1.03$ for the 1976-2006 data (bottom).
}
\end{figure}

In Figure \ref{FIG_DJFULL} we show a fit of the Hull-White model in the BO approximation
for daily returns of the DJIA index. We find $\beta = 0.861$ and $\theta=1.03$.
One can see that the resulting Tsallis distribution provides a much better fit compared to
the usual Gaussian model for the log-returns. It is important to notice that the parameters for the
Tsallis distribution are directly related to the Hull-White stochastic volatility model, providing 
a microscopic origin for this non-Gaussian distribution. 


\vs
\section{Discussion and Conclusions}
It has been known for quite sometime that the distribution of returns are not
Gaussian, as assumed by the SFM. What has been the subject of some
debate in the literature is the nature of the processes that cause
this deviation from normality.
 
Our results have shown that the observed non-Gaussian, fat-tailed distribution 
of returns can be obtained microscopically by two coupled independent stochastic 
processes, one for the returns and the other for the volatility of the returns.
In our analysis we assumed that the stochastic volatility process 
quickly reaches a stationary distribution and acts
as a background for the return process. We then obtain the distribution of the returns 
by integrating over the instantaneous volatility that is distributed according
to the stationary distribution.  This is analogous to the Born-Oppenheimer approximation
in Physics and we showed that it leads in this case to satisfactory results.
In particular, we demonstrated that all models of stochastic volatility should
exhibit a scaling relation in the time lag of zero-drift modified log-returns and we verified that the Dow-Jones Industrial
Average index indeed follows this scaling from time lags of of 1 day up to  500 days, 
beyond which data becomes sparse.

We provided a microscopic explanation for a Tsallis distribution of log-returns.
The parameters of a microscopic Hull-White stochastic volatility model uniquely determine the 
parameters of the macroscopic Tsallis distribution. We estimated these parameters from a fit 
to 30 years Dow Jones index daily data, concluding that they do result in a much better agreement 
than the SMF Gaussian distribution. Since the scaling is observed
in the data, this daily fit determines the distribution for all other time lags.
The transition to Gaussian distributions for large time lags occurs because the fat tails are 
pushed away as the distribution broadens with time and the central region of the distribution 
is well described by a Gaussian distribution. 
This can be seen by expanding eq.(\ref{PX_TSALLIS}) for large time lags, 
a well known fact in the literature \cite{SilvaYakovenko}.

In a more general approach one should attemp to reconstruct directly $a(v)$ and $b(v)$ from data, 
which is tantamount to reconstruct the best stochastic volatility model compatible with data. 
This attempt is analogous to deciphering the Planck distribution and 
hence the temperature for distant stars from their radiation spectra.
In this respect the BO approximation used in the Hull-White model should be 
viewed as yet another particular model and as such it should be tested. 
This was pursued in this paper.

\section*{Acknowledgments}
We would like to thank J\'anos Kert\'esz and especially 
Victor Yakovenko for a very careful reading of the paper and
many discussions about the assumptions made.
R.~Rosenfeld thanks CNPq for partial financial support. 
T.~S.~Bir\'o acknowledges the warm hospitality of the IFT at UNESP, S\~ao Paulo, Brazil,
and the partial support by the Hungarian National Research Fund OTKA (T49466).

\end{document}